\documentclass[11pt,a4paper]{article}
\usepackage[hyperref]{eacl2021}
\usepackage{times}
\usepackage{latexsym}
\usepackage{enumitem}
\usepackage{amsmath}
\usepackage{graphicx}
\usepackage{multirow}
\usepackage{verbatim}
\usepackage{url}

\usepackage{microtype}

\usepackage{xcolor}
\newcommand{\jiatong}[1]{\textcolor{blue}{\bf\small [#1 --jiatong]}}

\newcommand{\kev}[1]{\textcolor{green}{\bf\small [#1 --Kev]}}

\aclfinalcopy 
\def\aclpaperid{433} 

\DeclareMathOperator*{\argmax}{arg\,max}

\title{
Leveraging End-to-End ASR for Endangered Language Documentation: An Empirical Study on Yolox\'ochitl Mixtec}

\author{Jiatong Shi$^1$ \quad Jonathan D. Amith$^2$ \quad Rey Castillo García$^3$\\
\textbf{Esteban Guadalupe Sierra \quad Kevin Duh$^1$ \quad Shinji Watanabe$^1$}\\
$^1$The Johns Hopkins University, Baltimore, Maryland, United States\\
$^2$Department of Anthropology, Gettysburg College\\
$^3$Secretaría de Educación Pública, Estado de Guerrero, Mexico\\
{\tt \{jiatong\_shi@, kevinduh@cs.\}jhu.edu} \\
{\tt \{jonamith, reyyoloxochitl, estebanyoloxochitl\}@gmail.com} \\
{\tt shinjiw@ieee.org}\\
}

\begin{document}

\maketitle
\begin{abstract}

``Transcription bottlenecks", created by a shortage of effective human transcribers are one of the main challenges to endangered language (EL) documentation. Automatic speech recognition (ASR) has been suggested as a tool to overcome such bottlenecks. Following this suggestion,
we investigated the effectiveness for EL documentation of end-to-end ASR, which unlike Hidden Markov Model ASR systems, eschews linguistic resources but is instead more dependent on large-data settings. 
We open source a Yolox\'ochitl Mixtec EL corpus. First, we review our method in building an end-to-end ASR system in a way that would be reproducible by the ASR community. We then propose a novice transcription correction task and demonstrate how ASR systems and novice transcribers can work together to improve EL documentation. We believe this combinatory methodology would mitigate the transcription bottleneck and transcriber shortage that hinders EL documentation.
\end{abstract}

\section{Introduction}
\label{sec: intro}

\citet{grenoble2011handbook} warned that half of the world's 7,000 languages would disappear by the end of the 21st century. Consequently, a concern with endangered language documentation has emerged from the convergence of interests of two major groups: (1) native speakers who wish to document their language and cultural knowledge for future generations; (2) linguists who wish to document endangered languages to explore linguistic structures that may soon disappear. Endangered language (EL) documentation aims to mitigate these concerns by developing and archiving corpora, lexicons, and grammars \citep{lehmann1999documentation}. 
There are two major challenges: 


\paragraph{(a) Transcription Bottleneck:} 
The creation of EL resources through documentation is extremely challenging, primarily because the traditional method to preserve primary data is not simply with audio recordings but also through time-coded transcriptions. In a best-case scenario, texts are presented in interlinear format with aligned parses and glosses along with a free translation \citep{anastasopoulos2017case}. But interlinear transcriptions are difficult to produce in meaningful quantities: (1) ELs often lack a standardized orthography (if written at all); (2) invariably, few speakers can accurately transcribe recordings. Even a highly skilled native speaker or linguist will require a minimum of 30 to 50 hours to simply transcribe one hour of recording \citep{michaud2014towards, zahrer2020towards}. Additional time is needed for parse, gloss, and translation. This creates what has been called a ``transcription bottleneck", a situation in which the expert transcribers cannot keep up with the amount of recorded material for documentation. 

\paragraph{(b) Transcriber Shortage:} 
It is generally understood that any viable solution to the transcription bottleneck must involve native speaker transcribers. Yet usually few, if any, native speakers have the skills (or time) to transcribe their language. Training new transcribers is one solution, but it is time-consuming, especially with languages that present complicated phonology and morphology. The situation is distinct for major languages, for which transcription can be crowd-sourced to speakers with little need for specialized training \citep{das2016investigation}. In Yolox\'ochitl Mixtec (YM; Glottocode=yolo1241, ISO 639-3=xty), the focus of this study, training is time-consuming: after one-year part-time transcription training, a proficient native speaker, Esteban Guadalupe Sierra, still has problems with certain phones, particularly tones and glottal stops. Documentation requires accurate transcriptions, a goal yet beyond even the capability of an enthusiastic speaker with many months of training. 
\vspace{0.3cm}
As noted, ASR has been proposed to mitigate the Transcription Bottleneck and create increasingly extensive EL corpora. Previous studies first investigated HMM-based ASR for EL documentation \citep{cavar2016endangered, mitra2016automatic, adams2018evaluating, jimerson2018improving, jimerson2018asr, michaud2018integrating, cruz2019deploying, thai2020fully, zahrer2020towards, gupta-boulianne-2020-automatic}. Along with HMM-based ASR, natural language processing and semi-supervised learning have been suggested as a way to produce morphological and syntactic analyses. As HMM-based systems have become more precise, they have been increasingly promoted as a mechanism to bypass the transcription bottleneck. However, ASR's context for ELs is quite distinct from that of major languages. Endangered languages seldom have sufficient extant language lexicons to train an HMM system and invariably suffer from a dearth of skilled transcribers to create these necessary resources \citep{gupta2020speech}.

As we have confirmed with this present study, end-to-end ASR systems have shown comparable or better results over conventional HMM-based methods \citep{graves2014towards, chiu2018state, pham2019very, karita2019comparative}. As end-to-end systems directly predict textual units from acoustic information, they save much effort on lexicon construction. Nevertheless, end-to-end ASR systems still suffer from the limitation of training data. Attempts with resource-scarce languages have relatively high character (CER) or word (WER) error rates \citep{thai2020fully, matsuura2020speech, hjortnaes2020towards}. It has nevertheless become possible to utilize ASR with ELs to reduce significantly, but not eliminate, the need for human input and annotation to create acceptable (``archival quality") transcriptions.

\paragraph{This Work:}
This work represents end-to-end ASR efforts on Yolox\'ochitl Mixtec (YM), an endangered language from western Mexico. The YMC\footnote{Specifically, we used material from the community of Yoloxóchitl (YMC), one of four in which YM is spoken.} corpus comprises two sub-corpora. The first (``YMC-EXP", expert transcribed, corpus) includes 100 hours of transcribed speech that have been carefully checked for accuracy. 
We built a recipe of the ESPNet \citep{watanabe2018espnet} that shows the whole process of constructing an end-to-end ASR system using the YMC-EXP corpus.\footnote{\url{https://github.com/espnet/espnet/tree/master/egs/yoloxochitl_mixtec/asr1}} 
The second corpus, (``YMC-NT", native trainee, corpus) includes 8+ hours of additional recordings not included in the YMC-EXP corpus. This second corpus contains novice transcriptions with subsequent expert corrections that has allowed us to evaluate the skill level of the novice. Both the YMC-EXP and YMC-NT corpora are publicly available at OpenSLR under a CC BY-SA-NC 3.0 License.\footnote{\url{http://www.openslr.org/89/}}

The contributions of our research are:
\begin{itemize}
    \item A new Yolox\'ochitl Mixtec corpus to support ASR efforts in EL documentation.
    \item A reproducible workflow to build an end-to-end ASR system for EL documentation.
    \item A comparative study between HMM-based ASR and end-to-end ASR, demonstrating the feasibility of the latter. To test the framework's generalizability, we also experiment with another EL: Highland Puebla Nahuat (Glottocode=high1278; ISO 639-3=azz).
    \item An in-depth analysis of errors in novice transcription and ASR. Considering the discrepancies in error types, we propose Novice Transcription Correction (NTC) as a task for the EL documentation community. A rule-based method and a voting-based method are proposed.\footnote{A system combination method, Recognizer Output Voting Error Reduction \citep{fiscus1997post})} In clean speech, the best system reduces relative word error rate in the novice transcription by 38.9\% .
    
\end{itemize}

\section{Corpus Description}
\label{sec: corpus-description}


In this section, we first introduce the linguistic specifics for YM and YMC. Then we discuss the recording settings. Since YM is a spoken language without a standardized textual format, we next explain the transcription style designed for this language. Finally, we offer the corpus partition and some statistics regarding corpora size.

\subsection{Linguistic Specifics for Yolox\'ochitl Mixtec}

Yolox\'ochitl Mixtec is an endangered, relatively low-resource Mixtecan language. 
It is mainly spoken in the municipality of San Luis Acatlán, state of Guerrero, Mexico. It is one of some 50 languages in the Mixtec language family, which is part of a larger unit, Otomanguean, that \citet{suarez1983mesoamerican} considers ``a `hyper-family' or `stock'." Mixtec languages (spoken in Oaxaca, Guerrero, and Puebla)  are highly varied, resulting from approximately 2,000 years of diversification. 

YM is spoken in four communities: Yolox\'ochitl, Cuanacaxtitlan, Arroyo Cumiapa, and Buena Vista. Mutual intelligibility among the four YM communities is high despite significant differences in phonology, morphology, and syntax. All villages have a simple segmental inventory but significant though still undocumented variation in tonal phonology. YMC (refering only to the Mixtec of the community of Yolox\'ochitl [16.81602, -98.68597]) manifests 28 distinct tonal patterns on 1,451 identified bimoraic lexical stems. 
The tonal patterns carry a significant functional load in regards to the lexicon and inflection. For example, 24 distinct tonal patterns on the bimoraic segmental sequence [nama] yield 30 words (including six homophones). This ample tonal inventory presents challenges to both a native speaker learning to write and an ASR system learning to recognize. Notably, it also introduces difficulties in constructing a language lexicon for training HMM-based systems.

\subsection{Recording Settings}
\label{ssec: recording settings}
There are two corpora used in this study. The first (YMC-EXP) was used for ASR training. The second (YMC-NT) was used to train the novice speaker (e.g., set up a curriculum for him to learn how to transcribe) and for Novice Transcription Correction. The YMC-EXP corpus comprises expert transcriptions used as the gold-standard reference for ASR development. The YMC-NT corpus has paired novice-expert transcription as it was used to train and evaluate the novice writer.

The corpus used for ASR development comprises mostly conversational speech in two-channel recordings (split for training). Each conversation is with two speakers and each of the two speakers was fitted with a separate head-worn mic (usually a Shure SM10a). 
Over two dozen speakers (mostly male) contributed to the corpus. The topics and their distribution were varied (plants, animals, hunting/fishing, food preparation, ritual speech). 
The YMC-NT corpus comprises single-channel field recordings made with a Zoom H4n at the moment plants were collected during ethnobotanical research. Speakers were interviewed one after another; there is no overlap. However, the recordings often registered background sounds (crickets, birds) that we expected would negatively impact ASR accuracy more than seems to have occurred. The topic was always a discussion of plant knowledge (a theme of only 9\% of the YMC-EXP corpus). Expectedly, there were many out-of-vocabulary (OOV) words (e.g., plant names not elsewhere recorded) in this YMC-NT corpus.\footnote{After separating enclitics and prefixes as separate tokens, the OOV rate in YMC-NT is 4.84\%.}

\subsection{Corpus Transcription}
\label{ssec: corpus transcription}

\paragraph{(a) Transcription Level:} 
The YMC-EXP corpus presently has two levels of transcription: (1) a practical orthography that represents underlying forms; (2) surface forms.
The underlying form marks prefixes (separated from the stem by a hyphen), enclitics (separated by an = sign), and tone elision (with the elided tones in parentheses). All these ``breaks" and phonological processes disappear in the surface form. For example, the underlying $be'^3e^3$=$an^4$ (house=3sgFem; 'her house') surfaces as $be'^3\tilde{a}^4$. And $be'^3e^{(3)}$=$^2$ ('my house') surfaces as $be'^3e^2$.  Another example is the completive prefix $ni^1$-, which is separated from the stem as in $ni^1$-$xi^3xi^{(3)}$=$^2$ (completive-eat-1sgS; 'I ate'). The surface form would be written $n\tilde{i}^1xi^3xi^2$. Again, processes such as nasalization, vowel harmony, palatalization, and labialization are not represented in the practical (underlying) orthography but are generated in the surface forms. The only phonological process encoded in the underlying orthography is tone elision, for which parentheses are used.

The practical, underlying orthography mentioned above was chosen as the default system for ASR training for three reasons: (1) it is easier than a surface representation for native speakers to write; (2) it represents morphological boundaries and thus serves to teach native speakers the morphology of their language; and (3) for a researcher interested in generating concordances for a corpus-based lexicographic project it is much easier to discover the root for `house' in $be'^3e^3$=$an^4$ and $be'^3e^{(3)}$=$^2$ than in the surface forms $be'^3\tilde{a}^4$ and $be'^3e^2$. 

\paragraph{(b) ``Code-Switching" in YMC:} 
Endangered, colonialized Indigenous languages often manifest extensive lexical input from a dominant Western language, and speakers often talk with ``code-switching" (for lack of a better term). Yolox\'ochitl Mixtec is no exception. Amith considered how to write such forms best and decided that Spanish-origin words would be written in Spanish and without tone when their phonology and meaning are close to that of Spanish. So Spanish \textit{docena} appears over a dozen times in the corpus and is written \textit{tucena}; it always has the meaning of `dozen'. All month and day names are also written without tones. Note, however, that Spanish \textit{camposanto} (`cemetery') is also found in the corpus and pronounced as $pa^3san^4tu^2$. The decision was made to write this with tone markings as it is significantly different in pronunciation from the Spanish origin word. In effect, words like $pa^3san^4tu^2$ are considered loans into YM and are treated orthographically as Mixtec. Words such as \textit{tucena} are considered ``code-switching" and written without tones.

\paragraph{(c) Transcription Process:} 
The initial time-aligned transcriptions were made in Transcriber \citep{barras1998transcriber}. However, given that Transcriber cannot handle multiple tiers (e.g., transcription and translation, or underlying and surface orthographies), the Transcriber transcriptions were then imported into ELAN \citep{wittenburg2006elan} for further processing (e.g., correction, surface-form generation, translation).

\begin{table}
\centering
\begin{tabular}{lllll}
\hline \textbf{Corpus} & \textbf{Subset} & \textbf{UttNum} & \textbf{Dur (h)}  \\ \hline
\multirow{3}{*}{\textbf{EXP}}  & Train  & 52763 & 92.46 \\
& Validation & 2470 & 4.01 \\
& Test & 1577 & 2.52 \\
\hline
\multirow{3}{*}{\textbf{EXP(-CS)}} & Train & 35144 & 58.60 \\
& Validation & 1301 & 2.16 \\
& Test & 2603 & 4.35 \\
\hline
\multirow{3}{*}{\textbf{NT}} & Clean-Dev & 2523 & 3.45 \\
& Clean-Test & 2346 & 3.31 \\
& Noise-Test & 1335 & 1.60 \\
\hline

\end{tabular}
\caption{\label{tab: YMC-corpus Partition} YMC Corpus Partition for EXP (corpus with expert transcription), EXP(-CS) (subset of EXP without ``code-switching"), NT (corpus with paired novice and expert transcription)}
\end{table}

\subsection{Corpus Size and Partition}
\label{ssec: corpus partition}
Though endangered, YMC does not suffer from the same level of resource limitations that affect most ASR work with ELs \citep{cavar2016endangered, jimerson2018improving, thai2020fully}. The YMC-EXP corpus, developed for over ten years, provided 100 hours for the ASR training, validation, and test corpora. There are 505 recordings from 34 speakers in the YMC-EXP corpus, and the transcription for the YMC-EXP were all carefully proofed by an expert native-speaker linguist. As shown in Table \ref{tab: YMC-corpus Partition}, we offer a train-valid-test split where there is no overlap in content between the sets. The partition considers the balance between speakers and relative size for each part.

As introduced in Section \ref{ssec: recording settings}, the YMC-NT corpus has \textit{both} expert and novice transcription. It includes only three speakers for a total of 8.36 hours. In the recordings of two consultants, the environment is relatively clean and free of background noise. The speech of the other individual, however, is frequently affected by background noise. This seems coincidental as all three were recorded together, one after the other in random order. But given this situation, we split the corpus into three sets: clean-dev (speaker EGS), clean-test (speaker CTB), and noise-test (speaker FEF; see Table \ref{tab: YMC-corpus Partition}).

The ``code-switching" discussed in \ref{ssec: corpus transcription} (b) introduces different phonological representations and makes it difficult to train an HMM-based model using language lexicons. Therefore, previous work \citep{mitra2016automatic} using the HMM-based system for YMC did not consider phrases with ``code-switching". To compare our model with their results, we have used the same experimental corpus in our evaluation. Their corpus (YMC-EXP(-CS)), shown in Table \ref{tab: YMC-corpus Partition}, is a subset of the YMC-EXP; the YMC-EXP(-CS) corpus does not contain ``code-switching" phrases, i.e., phrases with words that were tagged as Spanish origin and transcribed without tone.

\section{ASR Experiments}
\label{sec: experiments}

\subsection{End-to-End ASR}
\label{ssec: end-to-end asr}
As ESPNet \citep{watanabe2018espnet} is widely used in open-source end-to-end ASR research, our end-to-end ASR systems are all constructed using ESPNet. 
For the encoder, we employed the conformer structure \citep{gulati2020conformer}, while for the decoder we used the transformer structure to condition the full context, following the work of \citet{karita2019improving}. The conformer architecture is a state-of-the-art innovation developed from the previous transformer-based encoding methods \citep{karita2019comparative, guo2020recent}. A comparison between the conformer and transformer encoders shows the value of applying state-of-the-art end-to-end ASR to ELs. 

\subsection{Experiments and Results}
As discussed above, our end-to-end model applied an encoder-decoder architecture with a conformer encoder and a transformer decoder. The architecture of the model follows  \citet{gulati2020conformer} while its configuration follows the aishell conformer recipe from ESPNet.\footnote{See Appendix for details about the model configuration.} The experiment is reproducible using ESPNet. 

As the end-to-end system models are based on word pieces, we adopted CER and WER as evaluation metrics. They help demonstrate the system performances at different levels of graininess. But because the HMM-based systems were decoding with a word-based lexicon, for comparison to HMM we only use the WER metric. To thoroughly examine the model, we conducted several comparative experiments, as discussed in continuation.



\paragraph{(a) Comparison with HMM-based Methods:}
We first compared our end-to-end method with the Deep Neural Network-Hidden Markov Model (DNN-HMM) methods proposed in Mitra el al. (2016). In this work, Gammatone Filterbanks (GFB), articulation, and pitch are configured for the DNN-HMM model. This baseline is a DNN-HMM model using Mel Filterbanks (MFB). In recent unpublished work, Kwon and Kathol develop a latest state-of-the-art CNN-HMM-based ASR model\footnote{See Appendix for details about the model configuration.} for YMC based on the lattice-free Maximum Mutual Information (LF-MMI) approach, also known as ``chain model" \citep{povey2016purely}. The experimental data of the above HMM-based models is YMC-EXP(-CS) discussed in Section \ref{ssec: corpus partition}. For the comparison, our end-to-end model adopted the same partition to ensure fair comparability with their results.

Table \ref{tab: asr result with HMM} shows the comparison between DNN-HMM systems and our end-to-end system on YMC-EXP(-CS). It indicates that even without an external language lexicon the end-to-end system significantly outperforms both the DNN-HMM baseline models and the CNN-HMM-based state-of-the-art model.

\begin{table}[]
\centering
\begin{tabular}{lll}
\hline \textbf{Model}  & \textbf{Feature} & \textbf{WER} \\ \hline
DNN-HMM & MFB & 36.9  \\
\multirow{2}{*}{DNN-HMM} & GFB + Articu. & \multirow{2}{*}{31.1} \\
& +Pitch & \\
CNN-HMM & \multirow{2}{*}{MFCC} & \multirow{2}{*}{19.1}  \\
(Chain) & & \\
E2E-Conformer & MFB + Pitch & \textbf{15.4} \\
\hline
\end{tabular}
\caption{\label{tab: asr result with HMM} Comparison between HMM-based Models and the End-to-End Conformer (E2E-Conformer) Model on YMC-EXP(-CS) that is a subset of the YMC-EXP \textbf{without} ``code-switching".}
\end{table}

In Section \ref{ssec: corpus transcription} (b), we note that ``code-switching" is invariably present in EL speech (e.g., YMC). Thus, ASR models built on "code-switching-free corpora (like YMC-EXP[-CS]) are not practical for real-world usage. However, a language lexicon is available only for the YMC-EXP(-CS) corpus so we cannot conduct HMM-based experiments with either YMC-EXP or YMC-NT corpora. 

\paragraph{(b) Comparison with Different End-to-End ASR Architectures:} 
We also conducted experiments comparing models with different encoders and decoders on the YMC-EXP corpus. For a Recurrent Neural Network-based (E2E-RNN) model, we followed the best hyper-parameter configuration, as discussed in \citet{zeyer2018improved}. For a Transformer-based (E2E-Transformer) model, the same configuration from \citet{karita2019improving} was adopted. Both models shared the same data preparation process as the E2E-Conformer model.

Table \ref{tab: asr result} compares different end-to-end ASR architectures on the YMC-EXP corpus.\footnote{The train set in YMC-EXP is significantly larger than that in YMC-EXP(-CS), the YMC-EXP corpus from which all lines containing a Spanish-origin word have been removed.} 
The E2E-Conformer obtained the best results, obtaining significant WER improvement as compared to the E2E-RNN and the E2E-Transformer models. The E2E-Conformer's WER on YMC-EXP(-CS) is slightly lower than that obtained for the whole YMC-EXP corpus, despite a significantly smaller training set in the YMC-EXP(-CS) corpus. Since the subset excludes Spanish words, ``code-switching" may well be a problem to consider in ASR for endangered languages such as YM. 

\begin{table}[]
\centering
\begin{tabular}{lcc}
\hline
\multirow{2}{*}{\textbf{Model}}  & \textbf{CER} & \textbf{WER} \\ 
& dev/test & dev/test \\
\hline
E2E-RNN & 9.2/9.3 & 19.1/19.2  \\
E2E-Transformer & 7.8/7.9 & 16.3/16.7  \\
E2E-Conformer & \textbf{7.7}/\textbf{7.7} & \textbf{16.0}/\textbf{16.1}  \\
\hline
\end{tabular}
\caption{\label{tab: asr result} End-to-End ASR Results on YMC-EXP (corpus \textbf{with} ``code-switching")}
\end{table}

\paragraph{(c) Comparison with Different Transcription Levels:} 
In addition to comparing model architectures, we compared the impact of transcription levels on the ASR model. E2E-Conformer models with the same configurations were trained using both the surface and the underlying transcription forms, which are discussed in Section \ref{ssec: corpus transcription}. We also trained separate RNN language models for fusion and unigram language models to extract word pieces for different transcription levels.

\begin{table}[]
\centering
\begin{tabular}{lcc}

\hline
\multirow{2}{*}{\textbf{Transcription Level}}  & \textbf{CER} & \textbf{WER} \\ 
& dev/test & dev/test \\
\hline

Surface & 8.0/\textbf{7.6} & 16.6/16.3  \\
Underlying & \textbf{7.7}/7.7 & \textbf{16.0}/\textbf{16.1}  \\
\hline
\end{tabular}
\caption{\label{tab: asr result for transcription level} E2E-Conformer Results for Two Transcription Levels (Underlying represents morphological divisions and underlying phonemes before the application of phonological rules; Surface is reflective of spoken forms and lacks morphological parsing)}
\end{table}

Table \ref{tab: asr result for transcription level} shows the E2E-Conformer results 
for both underlying and surface transcription levels. As introduced in Section \ref{ssec: corpus transcription}, the surface form reduces several linguistic and phonological processes compared to the underlying practical form. The results indicate that the end-to-end system is able to automatically infer those morphological and phonological processes and maintain a consistent low error rate.

\paragraph{(d) Comparison with Different Corpus Sizes:} 
As introduced in Section \ref{sec: intro}, most ELs are considered low-resource for ASR purposes. To measure the impact of resource availability on ASR accuracy we trained the E2E-Conformer model on 10, 20, and 50 hours subsets of YMC-EXP. The results demonstrate the model performances over different sizes of resources.

\begin{table}[]
\centering
\begin{tabular}{lcc}
\hline
\multirow{2}{*}{\textbf{Corpus}}  & \textbf{CER} & \textbf{WER} \\ 
& dev/test & dev/test \\
\hline
10h & 19.4/19.5 & 39.1/39.2 \\
20h & 12.6/12.7 & 26.2/26.2 \\
50h & 8.6/8.7 & 18.0/18.0 \\
Whole (92h) & \textbf{7.7}/\textbf{7.7} & \textbf{16.0}/\textbf{16.1}  \\
\hline
\end{tabular}
\caption{\label{tab: asr result on different size} E2E-Conformer Results on Different Corpus Size}
\end{table}

Table \ref{tab: asr result on different size} shows the E2E-Conformer performances on different amounts of training data. It demonstrates how the model consumes data. As corpus size is incrementally increased, WER decreases significantly. It is apparent that the model still has the capacity to improve performance with more data. The result also indicates that our system can get reasonable performances from 50 hours of data. This would be an important guideline when we collect a new EL database.

\paragraph{(e) The Framework Generalizability:} 
To test the end-to-end ASR systems' generalization ability, we conducted the same end-to-end training and test procedures on another endangered language: Highland Puebla Nahuatl (high1278; azz). This corpus is also open access under the same CC license.\footnote{\url{http://openslr.org/92}} It comprises 954 recordings that total 185 hours 22 minutes, including 120 hours transcribed data in ELAN and 65 hours still only in Transcriber and not used in ASR training.\footnote{The recordings are almost all with two channels and two speakers in natural conversation.}

Table~\ref{tab: asr result on Nahuatl} shows the performance of three different end-to-end ASR architectures on Highland Puebla Nahuatl. For this language the E2E-Conformer again offers better performances over the other models. Table~\ref{tab: asr result on different size-Nahuatl} shows the E2E-Conformer performances on different amounts of training data for Highland Puebla Nahuatl. We can observe that 50-hour is a reasonable size for an EL, which is similar to the experiments in Table~\ref{tab: asr result on different size}.
These experiments indicate the general ability to consistently apply end-to-end ASR systems across ELs. 

\begin{table}[]
\centering
\begin{tabular}{lcc}
\hline
\multirow{2}{*}{\textbf{Model}}  & \textbf{CER} & \textbf{WER} \\ 
& dev/test & dev/test \\
\hline
E2E-RNN & 10.3/9.9 & 26.8/25.4 \\
E2E-Transformer & \textbf{9.1}/9.1 & 23.7/\textbf{21.7} \\
E2E-Conformer & 9.9/\textbf{8.6} & \textbf{23.5}/\textbf{21.7} \\
\hline
\end{tabular}
\caption{\label{tab: asr result on Nahuatl} E2E-Conformer Results on another EL: Highland Puebla Nahuatl}
\end{table}

\begin{table}[]
\centering
\begin{tabular}{lcc}
\hline
\multirow{2}{*}{\textbf{Corpus}}  & \textbf{CER} & \textbf{WER} \\ 
& dev/test & dev/test \\
\hline
10h & 18.3/17.5 & 44.7/43.3 \\
20h & 14.2/12.9 & 34.8/33.3 \\
50h & 11.0/10.2 & 27.0/24.9 \\
Whole (120h) & \textbf{9.9}/\textbf{8.6} & \textbf{23.5}/\textbf{21.7}  \\
\hline
\end{tabular}
\caption{\label{tab: asr result on different size-Nahuatl} E2E-Conformer Results on another EL: High-land Puebla Nahuatl (Different Corpus Size)}
\end{table}


\section{Novice Transcription Correction}
\label{sec: novice transcription correction}

Finally, this paper presents novice transcription correction (NTC) as a task for EL documentation. That is, in this experiment we explore not only the possibility of using ASR to enhance the accuracy of a YM novice transcription but to combine both novice transcription and ASR to achieve accurate results that surpass that of either component. Below we first analyze patterns manifested in novice transcriptions. Next, we introduce two baselines that fuse ASR hypotheses and novice transcription for the NTC task.


\subsection{Novice Transcription Error}

As mentioned in Section \ref{sec: intro}, transcriber shortages have been a severe challenge for EL documentation. Before 2019, only the native speaker linguist, Rey Castillo García, could accurately transcribe the segments and tones of YMC. To mitigate the YMC transcriber shortage, in 2019 Castillo began to train another speaker, Esteban Guadalupe Sierra. First, a computer course was designed to incrementally teach Guadalupe segmental and tonal phonology.
In the next stage, he was given YMC-NT corpus recordings to transcribe. Compared to the paired expert transcription, the novice achieved a CER of 6.0\% on clean-dev, defined in Table \ref{tab: YMC-corpus Partition}.
However, it is not feasible to spend many months training speakers with no literacy skills to acquire the transcription proficiency achieved by Guadalupe in our project. Moreover, even with a 6.0\% CER, there are still enough errors so as to require significant annotation/correction by the expert, Castillo. The state-of-the-art ASR system (e.g., E2E-Conformer) shown in Table \ref{tab: asr result} gets an 8.2\% CER on the clean-dev set, more errors than the novice CER. So for YMC, ASR is still not a good enough substitute for a proficient novice.

\begin{table}[t]
\centering
\begin{tabular}{lll}
\hline \textbf{Error Types}  & \textbf{Novice} & \textbf{ASR} \\ \hline
Enclitics (=) & \textbf{96} & 243  \\
Prefixes (-) & 141 & \textbf{62}  \\
Glottal Stop (') & 341 & \textbf{209} \\
Parenthesis & 1607 & \textbf{302} \\
Tone & 4144 & \textbf{3241} \\
Stem-Nasal (n) & \textbf{0} & 6 \\
Others & \textbf{4263} & 10175 \\
\hline
Total & 10592 & 14232 \\
\hline
\end{tabular}
\caption{\label{tab: novice vs ASR} Character Error-type Distribution of Novice and ASR (by number of errors)}
\end{table}

\begin{figure}[t]
  \centering
  \includegraphics[width=0.7\linewidth]{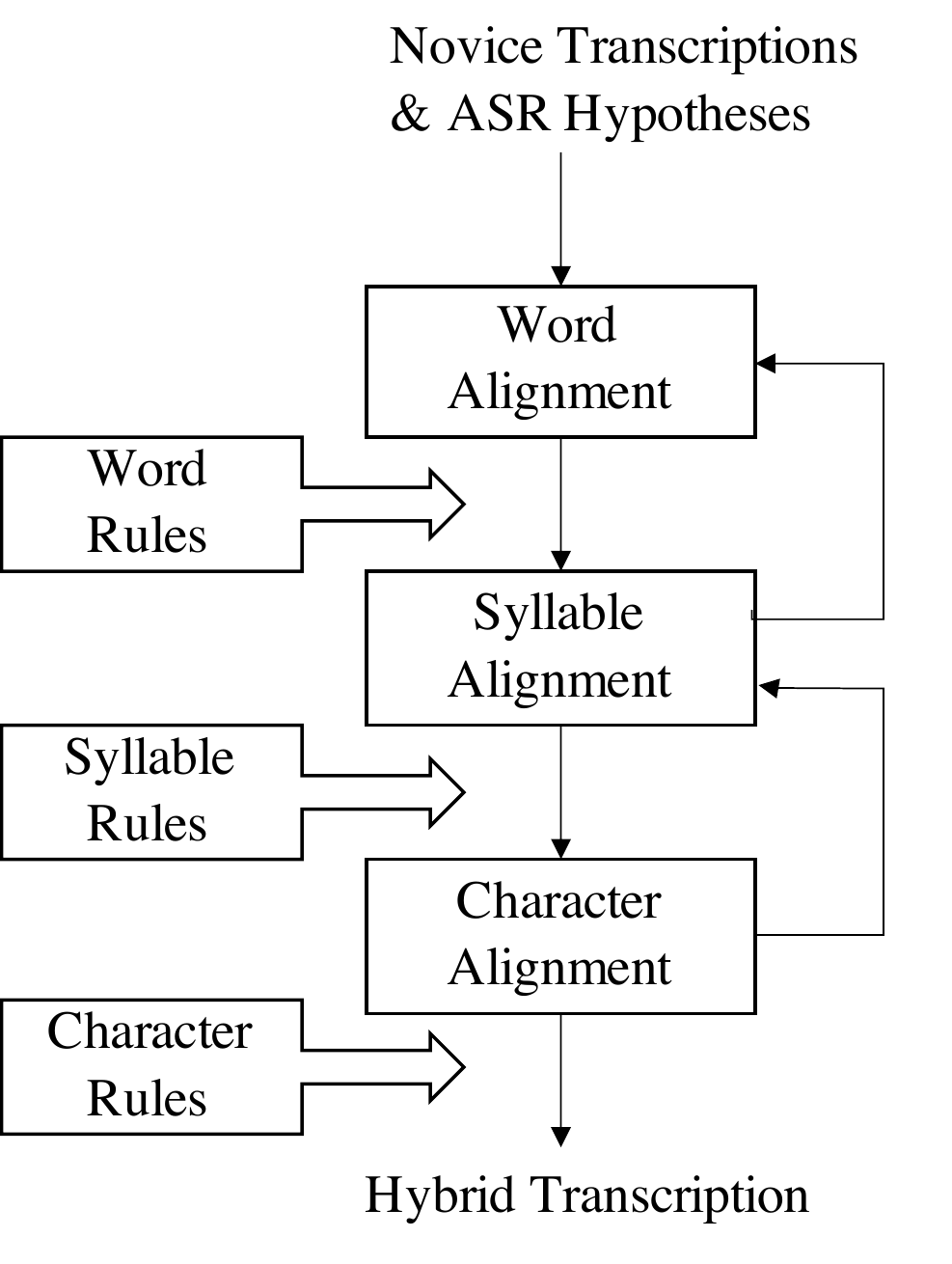}
  \caption{Novice-ASR Fusion Process}
  \label{fig:framework}
\end{figure}

As Amith and Castillo worked with the novice, they saw a repetition of types of errors that they worked to correct by giving the novice exercises focused on these transcription shortcomings. The end-to-end ASR, however, has demonstrated a different pattern of errors. For example, it developed a fair understanding of the rules for suppleting tones, marked by parentheses around the suppleted tones. 
Rather than over-specify the NTC correction algorithm, we first analyzed the error-type distribution using the Clean-dev from the YMC-NT corpus, as shown in Table \ref{tab: novice vs ASR}.

\begin{table*}[!tbh]
\centering
\begin{tabular}{lllllllll}
\hline
\multirow{2}{*}{\textbf{Model}}  & \multicolumn{2}{c}{\textbf{Clean-Dev}} & \multicolumn{2}{c}{\textbf{Clean-Test}} & \multicolumn{2}{c}{\textbf{Noise-Test}} &
\multicolumn{2}{c}{\textbf{Overall}}\\ 
& \textbf{CER} & \textbf{WER} & \textbf{CER} & \textbf{WER} & \textbf{CER} & \textbf{WER} & \textbf{CER} & \textbf{WER} \\
\hline
A. Novice & 6.0 & 21.5 & 6.4 & 22.6 & \textbf{8.4} & \textbf{26.6} & 6.8 & 23.1 \\
B. E2E-Transformer & 9.8 & 23.1 & 8.8 & 21.2 & 24.3 & 47.0 & 12.9 & 28.1 \\
C. E2E-Conformer & 8.2 & 19.6 & 8.2 & 19.1 & 23.6 & 44.1 & 12.0 & 25.3 \\
D. E2E-Conformer(50h) & 10.5 & 25.0 & 9.9 & 23.7 & 25.7 & 50.1 & 14.0 & 30.5 \\
\hline
E. Fusion1 (A+C) & 6.3 & 20.6 & 6.9 & 22.0 & 13.1 & 38.6 & 8.2 & 25.4 \\
F. Fusion1 (A+D) & 7.0 & 22.9 & 7.5 & 24.5 & 14.0 & 41.5 & 8.9 & 28.0 \\
G. Fusion2 (A+C) & 5.1 & 17.6 & 5.5 & 18.7 & 9.6 & 30.3 & 6.3 & 21.1 \\
H. Fusion2 (A+D) & 5.5 & 19.4 & 5.9 & 20.4 & 10.1 & 32.6 & 6.8 & 23.0 \\
\hline
I. ROVER (A+B+C) & 4.7 & \textbf{14.6} & \textbf{4.6} & \textbf{13.8} & 12.4 & 32.6 & 6.5 & \textbf{18.5} \\
J. ROVER-Fusion2 (A+B+C+E) & \textbf{4.5} & 16.1 & 4.7 & 16.7 & 9.0 & 28.3 & \textbf{5.7} & 19.3 \\
\hline
\end{tabular}
\caption{\label{tab: ntc result} NTC Results on YMC-NT (the results are evaluated using the expert transcription in YMC-NT). Model D is trained with a 50-hour subset of the YMC-EXP as shown in Table~\ref{tab: asr result on different size}.}
\end{table*}

\subsection{Novice-ASR Fusion}
\label{ssec: novice-asr-fusion}
Rapid comparison of the types of errors for each transcription (novice and ASR) demonstrated consistent patterns and has led us to hypothesize that a fusion system might automatically correct many of these errors. Two baseline methods are examined for the fusion: a voting-based system \citep{fiscus1997post} and a rule-based system. 

The voting-based system follows the definition in \citep{fiscus1997post} that combines hypotheses from different ASR models with novice transcription.

The framework of rule-based fusion is shown in Figure \ref{fig:framework}. The rules are defined in different linguistic units: words, syllables, and characters. They assume a hierarchical alignment between the novice transcription and ASR hypotheses. The rules are applied to the transcription from word to syllable to character level. The rules are developed based on continual evaluation of the novice's progress. Thus they will be different but discoverable when applied to a new language.  
However, the general principle should be applicable to other ELs: Novice trainees will learn certain transcription tasks easier than others. Below we explain the rules for YMC.

\noindent{\textbf{Word Rules}}: If a word from the novice transcription is Spanish (i.e., no tones and no linguistic indications [-, =, '] that mark it as Mixtec), keep the novice transcription. If the novice has extra words, not in the ASR hypothesis, keep those extra words.

\noindent{\textbf{Syllable Rules}}: If a novice syllable is tone initial, use the corresponding ASR syllable. If the novice and the ASR have identical segments but different tones, use the ASR tones. When an ASR syllable has CVV or CV'V, and its corresponding novice syllable has CV,\footnote{A CV syllable can occur in a monomoraic word. But novice will often write a CV word when it should be CVV or CV'V. Stem-final syllables can be CV, CVV or CV'V. But novice tends to write CV in these cases.} use the ASR syllable (CVV or CV'V). If the tone from either transcription system follows a consonant (except a stem-final $n$), use the other system's transcription.

\noindent{\textbf{Character Rules}}: If the ASR has a hyphen, equal sign, parentheses, glottal stop which is absent from the novice transcription, then always trust the ASR and maintain the aforementioned symbols in the final transcription.

We apply the edit distance \citep{wagner1974string} to find the alignment between the ASR model hypothesis $\{C_1, ..., C_n\}$ and the Novice transcription $\{C'_1, ..., C'_m\}$. The $L_{I}$, $L_{D}$, $L_{S}$ are introduced in the dynamic function as the insertion, deletion, and substitution loss, respectively. In the naive setting, $L_{I}$, $L_{D}$ are both set to 1. The $L_{S}$ is set to 1 if $C_i$ is different from $C'_j$ and 0 otherwise. This setting is computation-efficient. However, it does not consider how the contents mismatch between the $C_i$ and $C_j$. Therefore, we adopt a hierarchical dynamic alignment. In this method, the character alignment follows the native setting. While the $L_{S}(C_i, C'_j)$ for syllable alignment is defined as the normalized character-level edit distance between $C_i$ and $C'_j$ as follows:
\begin{equation}
\begin{split}
    L_{S}(C_i, C'_j) = \frac{D[C_i, C'_j]}{ |C_i|}
\end{split}
\end{equation}
where the $|C_i|$ is the lengths of the syllable. Similarly, the $L_{S}(C_i, C'_j)$ for word alignment is defined based on syllable alignment.


\section{NTC Experiments}
\subsection{Experimental Settings}
The novice transcription, the E2E-Transformer model, and the E2E-Conformer model were considered as baselines for the NTC task. To evaluate the system for reduced training data, we also show our results of E2E-Conformer trained with a 50-hour subset. For the end-to-end models, we adopted the trained model from Section \ref{sec: experiments} with the same decoding set-ups. To test the effectiveness of the hierarchical dynamic alignment, we tested the data with two fusion systems, namely Fusion1 and Fusion2. The Fusion1 system used the naive settings of edit distance, while the Fusion2 system adopted the hierarchical dynamic alignment. Both fusion systems adopt rules defined in Section \ref{ssec: novice-asr-fusion}. Two configurations for voting-based methods were tested. The first ``ROVER" combined three hypotheses (i.e., the E2E-Transformer, the E2E-Conformer, and the Novice). In contrast, the ``ROVER-Fusion2" combined the Fusion2 system with the above three. 

\subsection{Results}
As shown in Table \ref{tab: ntc result}, voting-based methods and rule-based methods all significantly reduce the novice errors for clean speech.\footnote{Note that the rules are developed based on YM specifics, so the result cannot be applied to other languages directly. Readers should view it as a case study.} However, for the noise-test, the novice transcription is the most robust method.
For overall results, the ROVER system (model I) has a lower WER, while the ROVER-Fusion2 system (model J) reaches a lower CER. Model J significantly reduces specific errors, including tone errors (25\%), enclitic errors (50\%), and parentheses errors (87.5\%). In addition, models D, F, and H indicate that the system could still reduce clean-environment novice errors using ASR models trained with a 50-hour subset of the YMC-EXP corpus.

As we discussed in Section \ref{sec: novice transcription correction}, novice and ASR transcriptions manifest distinct patterns of error and thus can be used to complement each other. Table \ref{tab: ntc result} shows that our proposed rule-based and voting-based fusion methods can potentially eliminate the errors that come from the novice transcriber, and it can mitigate the transcriber shortage problems based on these fusion methods. However, we should note that a noisy recording condition would negatively affect a fusion approach as ASR does poorly under such conditions (\textgreater 23\% CER), and for practical purposes, the novice transcription alone (\textless 8.5\%) is much more accurate. In such conditions we should rely on the novice transcriber alone.

\section{Conclusion and Future Work}
\label{sec: conlusion}
This work presents an open-source endangered language corpus in Yolox\'ochitl Mixtec and a comparative and reproducible study on various approaches to end-to-end ASR. 
We demonstrate that end-to-end approaches are feasible and present comparable results over conventional HMM approaches, which require resources such as language lexicons not necessary with end-to-end ASR. 
Additionally, we propose novice transcription correction as a potential task for ASR in EL documentation. We examine two methods to approach this task. The first is a rule-based approach that uses hierarchical dynamic alignment and linguistic rules to perform novice-ASR hybridization. The second is a voting-based method that combines hypotheses from the novice and end-to-end ASR systems. Empirical studies on the YMC-NT corpus indicate that both methods significantly reduce the CER/WER of the novice transcription for clean speech. 

The above discussion suggests that a useful approach to EL documentation using both human and computational (ASR) resources might focus on training each system (human and ASR) for particular transcription tasks. 
If we know from the start that ASR will be used to correct novice transcriptions in areas of difficulty, we could train an ASR system to maximize accuracy in those areas that challenge novice learning.




\bibliography{eacl2021}
\bibliographystyle{acl_natbib}

\appendix

\section{Appendices}
\label{sec:appendix}

\paragraph{Experimental Settings for End-to-End ASR:} All the end-to-end ASR systems adopted the hybrid CTC/Attention architecture integrated with an RNN language model. The best model was selected on the basis of performance on the development set. The input acoustic features were 83-dimensional log-Mel filterbanks features with pitch features \citep{ghahremani2014pitch}. The window length and the frameshift were set to 25ms and 10ms. SpecAugmentation are adopted for data augmentation \citep{park2019specaugment}. The prediction targets were the 150-word pieces trained using unigram language modeling \citep{kudo2018sentencepiece} (both for surface and underlying form). All the end-to-end models are fused with RNN language models.\footnote{Our experiments show that the RNN language model reduces WER by about 1\%.} The CTC ratio for Hybrid CTC/Attention was set to 0.3. The decoding beam size was 20. Training and Testing are based on Pytorch.

\noindent{\textbf{E2E-Conformer Configuration}}: The E2E-Conformer used 12 encoder blocks and 6 decoder blocks. All the blocks adopted a 2048 dimension feed-forward layer and four-head multi-head-attention with 256 dimensions. Kernel size in the Conformer block was set to 15. For training, the batch size was set to 32. Adam optimizer with 1.0 learning rate and Noam scheduler with 25000 warmup-steps were used in training. We trained for a max epoch of 50. The parameter size is 43M.

\noindent{\textbf{E2E-RNN Configuration}}: The E2E-RNN used 3 encoder blocks and 2 decoder blocks. All the blocks adopt 1024 hidden units. Location-based attention adopted 1024-dim attention. Adadelta was chosen as the optimizer, and we trained for a max epoch of 15. The parameter size is 108M.

\noindent{\textbf{E2E-Transformer Configuration}}: The E2E-Transformer used 12 encoder blocks and 6 decoder blocks. All the blocks adopted a 2048 dimension feed-forward layer and four-head multi-head-attention with 256 dimensions. Adam optimizer with 1.0 learning rate and Noam scheduler with 25000 warmup-steps were used in training. We trained for a max epoch of 100. The parameter size is 27M.

\paragraph{Experimental Settings for HMM-based ASR:}
The acoustic feature input for this model is 40 dimensional Mel Frequency Cepstral Coefficients (MFCC). The lexicon for HMM-based models is phone-based. The transcriptions are mapped to surface representations and then to phones (a total of 197 phones, as each tone for a given vowel, is a different phone). There are 22,465 total entries in the lexicon. The chain model is trained with a sequence-level objective function and operates with an output frame rate of 30 ms, three times longer than the previous standard. The longer frame rate increases decoding speed, which in turn makes it possible to operate with a significantly deeper DNN architecture for acoustic modeling. The best results were achieved with a neural network based on the ResNet architecture \citep{szegedy2017inception}. This consists of an initial layer for Linear Discriminative Analysis (LDA) transformation and subsequent alternating 160-dimensional bottleneck layers, adding up to 45 layers in total. The DNN acoustic model is then compiled with a 4-gram language model into a weighted finite-state transducer for word sequence decoding.

\end{document}